\documentclass[prx,aps,twocolumn,showpacs,superscriptaddress,nofootinbib]{revtex4-1}

\usepackage{hyperref}
\usepackage{graphicx}
\usepackage{dcolumn}
\usepackage{bm}
\usepackage[bbgreekl]{mathbbol}
\usepackage{amsmath,dsfont}
\usepackage[normalem]{ulem}
\usepackage{color}
\usepackage{booktabs}
\usepackage{comment}
\usepackage{dblfloatfix}
\usepackage{romannum}

\begin{document}

\title{Bending Instability of Rod-shaped Bacteria}

\author{Luyi Qiu}
\address{John A. Paulson School of Engineering and Applied Sciences, Harvard University, Cambridge, MA 02138, USA}

\author{John W. Hutchinson}
\address{John A. Paulson School of Engineering and Applied Sciences, Harvard University, Cambridge, MA 02138, USA}

\author{Ariel Amir}
\address{John A. Paulson School of Engineering and Applied Sciences, Harvard University, Cambridge, MA 02138, USA}

\date{\today}

\begin{abstract}
  A thin-walled tube, e.g., a drinking straw, manifests an instability when bent by localizing the curvature change in a small region.
  This instability has been extensively studied since the seminal work of Brazier nearly a century ago.
  However, the scenario of pressurized tubes has received much less attention.
  Motivated by rod-shaped bacteria such as \textit{E. coli}, whose cell walls are much thinner than their radius and are subject to a substantial internal pressure,  we study, theoretically, how this instability is affected by this internal  pressure.
  In the parameter range relevant to the bacteria, we find that the internal pressure significantly postpones the onset of the instability, while the bending stiffness of the cell wall has almost no influence.
  This study suggests a new method to infer turgor pressure in rod-shaped bacteria from bending experiments.
\end{abstract}

\maketitle

\begin{figure}[h!]
    \centering
    \includegraphics[width=\columnwidth]{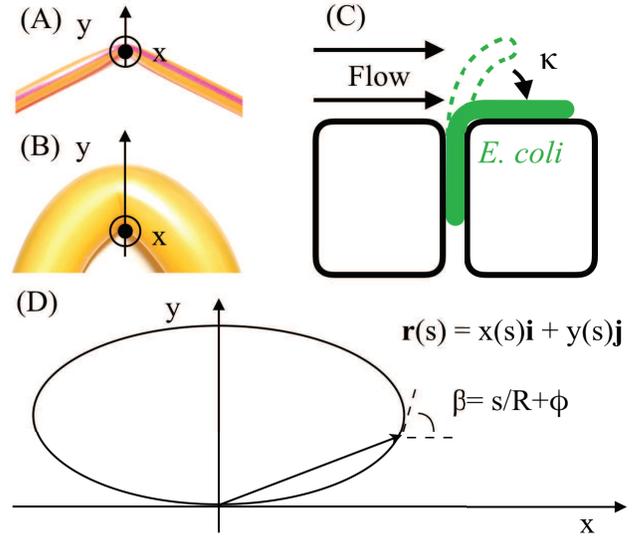}
    \caption{\textbf{Buckling of (A) a straw, (B) a birthday balloon and
    (C) an \textit{E. coli} cell in a mother machine. (D) Geometry notation.}
A deformed cross-section is shown here.
The points on the curve can be described by $\textbf{r}(s)=x(s)\textbf{i}+y(s)\textbf{j}$
as a vector pointing from the coordinate origin to it.
$s$ is the distance measured along the curve. $\mu$ is corresponding dimensionless parameter $\mu=s/R$.
$\beta$ is the angle between the $x$-axis and the vector tangent to the curve
in the deformed configuration.
$\phi(\mu)$ is the rotation angle, i.e., $\phi(\mu)=\beta-\mu$.
    }
    \label{fig:geometry}
\end{figure}

\textit{Introduction}.---As can be intuited from everyday experience, a thin-walled cylindrical tube such as a drinking straw
subject to bending reaches a critical curvature at which instability occurs, localizing most of the curvature change into a narrow region (Figure \ref{fig:geometry} (A) and (B)).
This instability has been extensively studied since the seminal work of Brazier nearly a century ago \cite{brazier1927}.
Brazier calculated, approximately, the external torque needed to bend the tube
to a given curvature of its long axis, and found that the dependence is non-monotonic with a maximum value.
Localization of the curvature change is expected at the curvature where the torque reaches a maximum.
This instability is characterized by its dependence on the geometry
rather than material nonlinearity.
Another candidate for instability of a thin-walled tube is the wrinkling effect.
As identified independently by Timoshenko \cite{timo1910}, Lorenz \cite{lorenz1908}
and Southwell \cite{southwell1914},
when the lateral compressive stress reaches a critical value
the system will develop periodic structures on the surface to minimize elastic energy. Under increasing overall curvature, the wrinkles grow and trigger localization of the overall curvature. An extensive study of the competition between wrinkling and the Brazier instability for thicker metal shells which undergo plastic deformation prior to experiencing bending instability has been given by Kyriakides and Corona in their book on buckling of undersea pipelines \cite{kyriakides2007}.

As the tube is bent, whenever the Brazier or the wrinkling instability is reached,
the stress localizes, resulting in the characteristic kinks shown in Figure \ref{fig:geometry} (A) and (B). It is not apparent a priori which of the two instabilities will occur first and this will be addressed. Further, we shall show that the structural instability can be used to infer the mechanical properties of the system, e.g. turgor pressure, for rod-shaped bacteria.

\textit{Model}.---
Here we consider a pressurized capped cylindrical tube of radius $R$, thickness $t$ and length $L \gg R$
with inner pressure larger than the external pressure by $p$.
We bend the pressurized shell to a longitudinal curvature $\kappa$ with no axial constraint.
Away from the capped ends, each cross-section behaves identically,
and we take one cross-section in the $(x, y)$ plane
as representative in Figure \ref{fig:geometry} (D).
The incremental material response measured from the cylindrical pressurized state is taken to be linear though the cylindrical swelling due to $p$ may involve nonlinear elastic deformation, depending on the constitutive model.
For rod-shaped bacteria such as \textit{E. coli}, elasticity of the cell wall is believed to be non-isotropic, presumably due to the preferential organization of the stiffer glycan strands in the circumferential direction and softer peptides along the longitudinal direction. Note also that the stresses on a pressurized cylinder are such that the circumferential stress is twice as large as the axial one, which for finite strains will also lead to non-isotropy.
Therefore, we use the general orthotropic relation for stress and strain increments in the axial and circumferential directions (SM \Romannum{1}).
For mathematical convenience and clarity, the deformation occurring during bending after $p$ has been applied is assumed to be inextensional in the circumferential direction.
This is similar to the hypothesis used by Euler in his Elastica framework.
The deformation occurring during bending is then fully characterized by the rotation function $\phi(\mu)$ defined in Figure \ref{fig:geometry} (D), the imposed curvature $\kappa$ of the axial line element lying along $y = 0$ and the axial strain change $\Delta \varepsilon_0$ of that line element.

Prior to bending, the resultant membrane stresses are $N_{\theta}^0=pR$ and $N_z^0=pR/2$ , and the bending moments in the tube wall are negligible. The change in the energy due to bending of the system under fixed $p$ includes the sum of the changes in bending energy, stretching energy and potential energy of pressure \cite{landau1986}.
Under the circumferential inextensibility assumption,
the circumferential membrane strain remains unchanged upon bending.
The contribution to the bending energy in the wall of the tube associated with axial curvature, $\Delta K_z \approx -\kappa$, is negligible compared to the axial stretching energy and is ignored. With $\Delta \Phi$ as the change in energy per unit length from the straight pressurized state, we have (SM \Romannum{1}):
\begin{equation}
  \begin{aligned}
   \Delta \Phi & = \int_0^{2\pi R}
   \left( N_z^0 \Delta \varepsilon_z
   + \frac{1}{2}\Delta N_z \Delta \varepsilon_z + \frac{1}{2}\Delta M_{\theta} \Delta K_{\theta} \right)ds \\
   & -p \Delta V \;.
  \end{aligned}
\end{equation}
\noindent
Here $\Delta \varepsilon_z = \Delta \varepsilon_0 + \kappa y$ is the change of axial strain, $\Delta K_{\theta}= d\phi/ds$ is the circumferential curvature change, $\Delta V$ is volume change per unit length, $\Delta N_z = S_z \Delta \varepsilon_z$ is the increment of resultant membrane stresses and $\Delta M_{\theta}=D_{\theta}\Delta K_{\theta}$ is the increment of shell wall bending moments.  $S_{\alpha}= E_{\alpha} t /(1-\nu_{\theta z}\nu_{z \theta})$ and $D_{\alpha}=S_{\alpha} t^2 /12$ ($\alpha$ can be $z$ or $\theta$) are determined by material elastic properties.
For imposed $\kappa$, $\Delta \Phi$ can be expressed in terms of $\Delta \varepsilon_0$ and $\phi(\mu)$ (SM \Romannum{1}):
\begin{equation}\label{eq:energy_expression}
  \begin{aligned}
    {\Delta \Phi} & = \frac{1}{2}\int_0^{2\pi}
    \left[\frac{D_{\theta}}{R^2}\left(\frac{d\phi}{d\mu}\right)^2 +S_z (\Delta \varepsilon_0+\kappa y)^2\right] R d\mu \\
    & +p \pi R^2\left[1+\Delta \varepsilon_0+\frac{\kappa}{2\pi}\int_0^{2\pi} yd\mu \right] \\
    & -p R ^2 \int_0^{2\pi} \frac{x}{R}(1+\Delta \varepsilon_0+\kappa y) \sin(\mu+\phi)d\mu \;. \\
  \end{aligned}
\end{equation}

\noindent For the convenience of calculation and presentation, it is useful to define and use  non-dimensional parameters.
The equations can be rendered dimensionless in multiple ways.
We can define a dimensionless geometry/material parameter in this system:
$\alpha = \sqrt{\frac{D_{\theta}}{S_zR^2}}=\sqrt{\frac{E_{\theta}}{12E_z}}\frac{t}{R}$.
Note that for thin tubes, $\alpha \ll 1$.
For \textit{E. coli} cells, $0.001 < \alpha < 0.01$ (SM \Romannum{2}). For thin shells, it is common to use the following dimensionless variables:
\begin{equation}
\label{eq:normalization_shell}
    \bar{\Phi}  = \frac{R}{D_{\theta}} \Delta \Phi \;,\;
    \bar{p}  = \frac{pR^3}{D_{\theta}} \;, \;
    \bar{\kappa}  = \frac{\kappa R}{\alpha} \;,\;
    \bar{M} = \frac{M}{\alpha S_z R^2}  \;,\;
\end{equation}
in which $M$ is the external torque needed to bend the tube.
The “shell” normalization is employed under the tacit assumption that $\bar{p}$ is of order unity \cite{calladine1983}. However, for \textit{E. coli}, $\bar{p}=O(10^4)$ (SM \Romannum{2}).
Consequently, in this pressure range it is more natural to use the following “balloon” normalization favoring the stretching stiffness:
\begin{equation}\label{eq:normalization_balloon}
  \hat{\Phi} = \frac{\Delta \Phi}{S_z R}\;, \;
  \hat{p} = \frac{pR}{S_z} \;, \;
  \hat{\kappa} = \kappa R \;, \;
  \hat{M} = \frac{M}{S_z R^2} \;.
\end{equation}
Note that $\bar{p}=\alpha^2 \hat{p}$ such that for the \textit{E. coli} cells $\hat{p}$ is of order unity (SM \Romannum{2}). In the results to follow we will illustrate both the “shell” and “balloon” normalizations. While the dimensionless quantities are different from one another, the form of the underlying governing equations is the same.

\noindent The state of the system for any imposed $\kappa$ can be determined by minimizing $\Delta\Phi$ with respect to the rotation function $\phi(\mu)$ and $\Delta \varepsilon_0$. $\Delta \varepsilon_0$ can also be determined by a force-balance equation (SM \Romannum{3}). To further proceed, $\phi(\mu)$ is discretized using a Fourier series representation.
Symmetry of the system about the y-axis requires $\phi(\mu)=-\phi(2\pi-\mu)$, and the boundary condition: $x(2\pi) = x(0) =0$ must be enforced. These lead to:
\begin{equation}\label{eq:phi_mu}
  \begin{aligned}
      \phi(\mu) & = \sum_{n=1}^{N} a_{n} \sin(n\mu) \;, \\
      0  & = \int_0^{2\pi} \cos(\mu +\phi(\mu)) d\mu \;. \\
  \end{aligned}
\end{equation}
For the special case of zero pressure, one can develop an approximate analytical solution for $a_n$ by expanding
the integrands of $x(\mu)=R \int_0^{\mu} \cos(\mu^\prime+\phi) d\mu^\prime$
and $y(\mu)=R \int_0^{\mu} \sin(\mu^\prime+\phi) d\mu^\prime$ using Taylor expansions of $\phi$ (SM \Romannum{4}). In the shell non-dimensionalization,
\begin{equation}
  a_n = -\frac{\bar{\kappa}^2}{4(n-1)^2n^2}a_{n-2} \;,\; a_1=0\;,\; a_2=-\frac{\bar{\kappa}^2}{8} \;.
\end{equation}
The dominant coefficient $a_2$ is much larger than all the others. This agrees with Brazier's result for zero pressure.

\noindent For non-zero positive pressure, we use the ansatz $\phi(\mu) = a_2 \sin(2\mu)$ as an approximation whose accuracy will be verified by numerical solutions shown later. Following the same Taylor
expansion approximation, we obtain (SM \Romannum{4}):
\begin{equation} \label{eq:a_2}
  a_2=-\frac{\bar{\kappa}^2}{8(1+\bar{p}/3)} = -\frac{\hat{\kappa}^2}{8(\alpha^2+\hat{p}/3)}\;,
\end{equation}
\begin{equation} \label{eq:delta_epsilon_0}
    \Delta \varepsilon_0 = -\hat{\kappa}\left(1+\frac{2}{3}a_2\right)-\frac{1}{3}\left(\hat{p}-\frac{4}{5}\hat{\kappa}\right)a_2^2+O(a_2^3) \;.
\end{equation}

\noindent
This allows us to compute the overall torque-curvature relation and the dependence of the maximum torque on the pressure. To compute the torque-curvature relation we use $M=\frac{\partial \Delta \Phi}{\partial \kappa}$ which gives for the two normalizations (SM \Romannum{5}):
\begin{equation}\label{eq:moment}
  \bar{M} = \pi \left[\bar{\kappa} -\frac{\bar{\kappa}^3}{8(1+\bar{p}/3)}  \right] \;\mbox{or}\;
  \hat{M} = \pi \left[\hat{\kappa} -\frac{\hat{\kappa}^3}{8(\alpha^2+\hat{p}/3)}  \right] \;.
\end{equation}
The critical curvature $\kappa_B$ for the Brazier instability occurs at the maximum of external torque $M$:
\begin{equation}\label{eq:k_B}
  \bar{\kappa}_B = \sqrt{\frac{8}{3}+\frac{8}{9}\bar{p}} \;\; \mbox{or} \;\;
  \hat{\kappa}_B = \sqrt{\frac{8}{3}\alpha^2+\frac{8}{9}\hat{p}} \;.
\end{equation}
Equation \ref{eq:k_B} reveals that the pressure can greatly postpone the onset of the Brazier instability. Interestingly, $a_2$ at the maximum torque is a constant $-1/3$ independent of material properties and the inner pressure.
In other words, the shape of the cross-section at the critical state is always the same for
the Brazier instability.
At the maximum torque, the tube cross-section is squeezed in the y-direction by about $22\%$ and its second moment is reduced by about $40\%$.

An accurate estimate of the onset of the wrinkling instability is obtained by making use of the formula for the axisymmetric buckling of a pressurized circular cylindrical shell of radius $\rho$ and thickness $t$ subject to a compressive axial stress $\sigma$. For a shell with the present incremental orthotropic properties,
the critical compressive stress $\sigma_c$ and the associated axial wavelength $l$ of the sinusoidal wrinkling mode are
\begin{equation}\label{eq:sigma_c}
    \sigma_c t = \frac{2}{\rho} \sqrt{D_zS_{\theta}}\;,\;\;
    l=2\pi \left(\frac{D_z}{S_{\theta}}\rho^2\right)^{1/4} \;.
\end{equation}
These formulas apply approximately to the wrinkling instability of the tube under bending if one identifies the critical stress with the maximum compressive stress in the ovalized tube, and $\rho$ is the circumferential radius of curvature at the position of maximum compression.
The validity of the approximation is because the wrinkling mode has a wavelength proportional to $\sqrt{\rho t}$ which is short compared to the radius of
the tube. Detailed calculations in the shell buckling literature \cite{seide1961buckling} have shown that the critical stress given by Equation \ref{eq:sigma_c} underestimates the local compressive stress at the onset of wrinkling in a thin elastic shell under bending by only a few percent. The thinner the shell, the more accurate the approximation.
In summary, the onset of the wrinkling instability is attained when the curvature $\kappa$ is sufficiently large so that $N_z^0+\Delta N_z = -\sigma_c t$ according to the critical stress in Equation \ref{eq:sigma_c}.
Employing the balloon normalization (Equation \ref{eq:normalization_balloon}) with the
expressions for $a_2$ (Equation \ref{eq:a_2}) and $\Delta \varepsilon_0$ (Equation \ref{eq:delta_epsilon_0}), one can obtain the following dimensionless equation for the overall curvature $\hat{\kappa}_w$ at the onset of wrinkling
(SM \Romannum{6}):
\begin{equation}\label{eq:k_w_original}
    8\left(\alpha^2+\frac{\hat{p}}{3}\right)
    \left(\frac{\hat{p}}{2}+2\alpha-\hat{\kappa}_w\right)
    +\left(\frac{2\hat{\kappa}_w}{3}-4\alpha\right)\hat{\kappa}_w^2=0 \;.
\end{equation}
with the associated torque given by Equation \ref{eq:moment}. Two special limits are worth noting.
If $\hat{p}=0$, Equation \ref{eq:k_w_original} becomes
\begin{equation} \label{eq:k_w_p_0}
    \bar{\kappa}_w^3 - 6\bar{\kappa}_w^2
    -12\bar{\kappa}_w +24 =0 \;.
\end{equation}
with the smallest positive solution $\bar{\kappa}_w=1.320$.
Thus, for the unpressurized tube, wrinkling occurs before the Brazier instability $\bar{\kappa}_B = \sqrt{8/3}=1.633$.
The other limit applies when the pressure is in the “balloon range”  and the tube is thin ($\alpha \ll 1$) such that
$\alpha$ is negligible in Equation \ref{eq:k_w_original}, leading to $\hat{\kappa}_w^3 - 4 \hat{p}\hat{\kappa}_w +2\hat{p}^2 =0$.
We therefore obtain $\hat{\kappa}_{w} = \frac{\hat{p}}{2} + O(\hat{p}^2)$.
\begin{figure}[!h]
    \centering
    \includegraphics[width=\columnwidth]{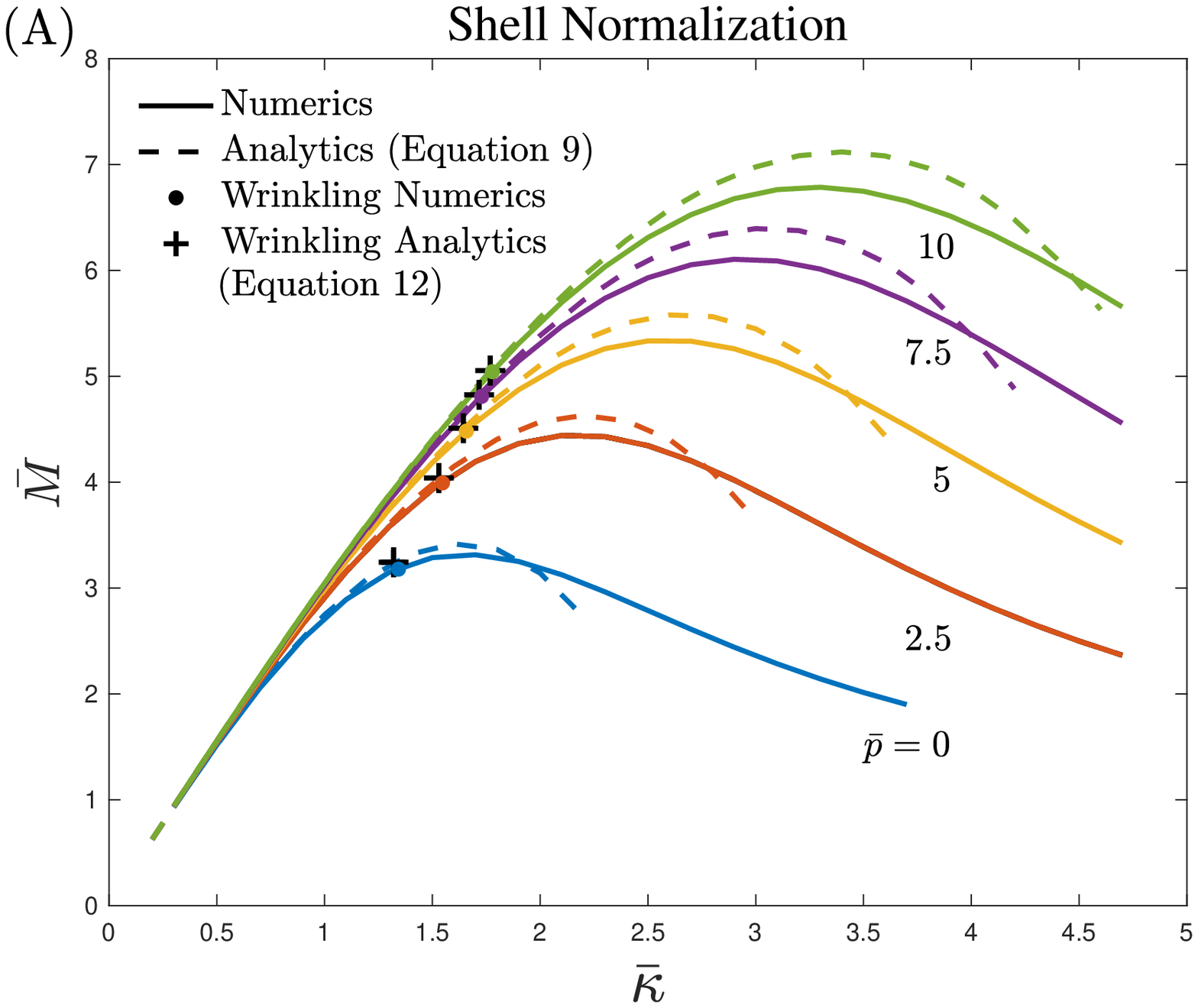}
    \includegraphics[width=\columnwidth]{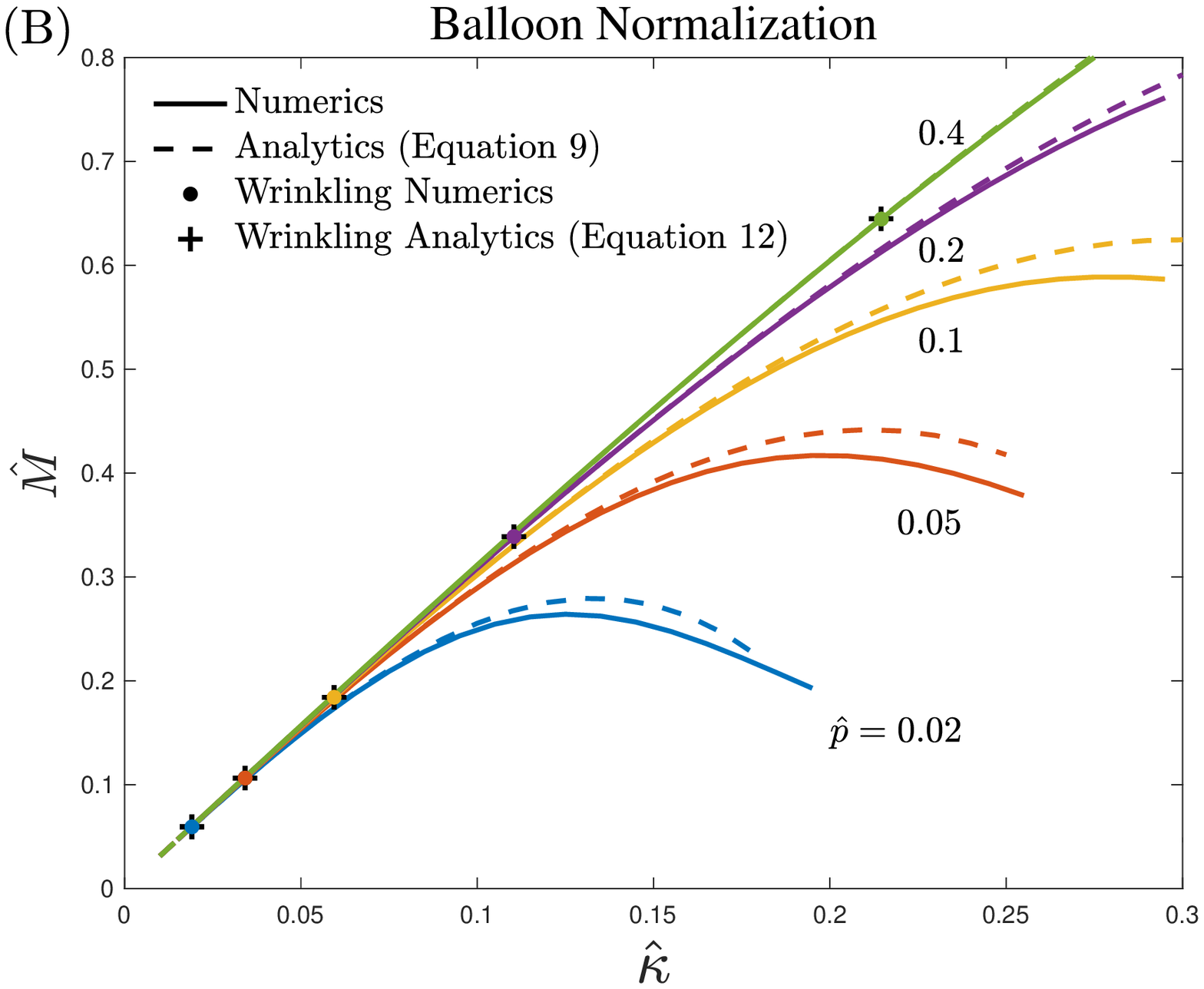}
    \caption{\textbf{Comparison of analytical and numerical results for torque-curvature relation and onset of wrinkling for typical elastic properties of \textit{E. coli} (SM \Romannum{2} Table 1, $\alpha \approx 0.005$).
    (A)} Shell normalization for low pressures. \textbf{(B)}  Balloon normalization for high pressures. The dashed lines are given by the formulas in Equation \ref{eq:moment},
    while the solid lines are based on numerical minimization of $\Delta \Phi$ (SM \Romannum{7}).
    The black cross is the onset of wrinkling predicted by Equation \ref{eq:k_w_original}. The color dots indicate the onset of wrinkling computed numerically.
    It is within $1\%$ of the numerical computed value.
    The critical curvatures for systems with different $\alpha$ are shown in SM \Romannum{8}.
    }
    \label{fig:M_k}
\end{figure}

\textit{Numerical results}.---
The torque-curvature relation (Equation \ref{eq:moment}) for the two normalizations is plotted in Figure \ref{fig:M_k} for various relevant dimensionless pressures where they can be compared with numerical results based on minimization of the energy functional $\Delta \Phi$ in Equation \ref{eq:energy_expression} (SM \Romannum{7}).
Included in Figure \ref{fig:M_k} on each of
the torque-curvature curves are solid dots marking the onset of the wrinkling instability computed numerically.
They agree well with the crosses, which are the wrinkling curvatures predicted (Equation \ref{eq:k_w_original}).
Over the entire range of pressures, in shell normalization or balloon normalization, wrinkling precedes attainment of the maximum moment, increasingly so as the pressure increases. Note that in the balloon regime wrinkling occurs on the initial linear segment of the torque-curvature curve for which $\hat{M} \approx \pi \hat{\kappa}$.
The validity of the incremental formulation is limited to relatively small incremental strains, not greater than $0.2$.
The maximum axial strains due to bending are of the order of $\kappa R$.
Note that for $\hat{p}= 0.4$ the critical wrinkling curvature is $\kappa_w R \approx 0.2$ . Thus, the wrinkling predictions are expected to be valid in the range $\hat{p}<0.4$.

\textit{A method to measure turgor pressure}.---
Our results can be utilized to provide a novel protocol for measuring turgor pressure in bacteria (Figure \ref{fig:geometry} (C) and SM \Romannum{9}), a task which has proved challenging over decades \cite{deng2011}. Previous works have shown that one may grow filamentous bacteria with large length to diameter ratios, and bend them either with optical tweezers \cite{wang2010} or with force generated by viscous drag due to fluid flow \cite{amir2014}. According to our results and for the relevant parameter range for \textit{E. coli}, as long as the wrinkling instability is not reached cell bending will be approximately linear in the force and independent of pressure, as validated experimentally \cite{amir2014}. Note, however, that here the incremental modulus $S_z$ can depend on $p$.
If the osmolarity of the media surrounding the cell is suddenly increased (by e.g. adding sugar) the turgor pressure drops abruptly \cite{rojas2014}, while the torque on the cell remains unperturbed. If the new turgor pressure is sufficiently low such that the wrinkling instability is reached -- as quantified by Equation \ref{eq:k_w_original} -- the cell would immediately buckle. Therefore by repeating this experiment for varying degrees of the hyperosmotic shock, the turgor pressure can be accurately measured. In fact, preliminary results using this protocol have shown it is feasible to achieve cell buckling upon osmotic shock for wild-type filamentous \textit{E. coli} \cite{private_communication}.
\emph{A priori} one might have envisioned that an alternative way to infer turgor pressure is from the $M-\kappa$ curve as indicated by Equation \ref{eq:moment}. However, for the parameters of \textit{E. coli}, the $M-\kappa$ curve shows barely any non-linearity before the instability point, as shown in Figure \ref{fig:M_k}.

\textit{Discussion}.---
In this work we revisited the long-standing problem of the Brazier effect, albeit for the understudied yet highly relevant scenario of a pressurized tube. While in structural mechanics applications the relevant pressure regimes are typically assumed to be associated with ``shells", where the pressure is sufficiently small in comparison with the bending rigidity, microbes such as bacteria are found to be in a qualitatively different ``balloon-like" regime with tremendous pressures outside the scope of previous theoretical work. By treating the problem using an Elastica framework we were able to obtain analytical formulas for the two potential instabilities that may arise when bending such highly pressurized tubes: one associated with a maximum in the torque-curvature relation, and the other associated with the onset of wrinkling at a critical compressive stress.
We corroborated our results numerically, finding good agreement between the approximate theory (assuming a particular mode of deformation dominates) and the precise numerics.

Within our theoretical approach, we found that the torque-curvature relations are well approximated by a linear dependence with a pressure-independent slope, and a cubic curvature contribution that strongly depends on the pressure as $\frac{1}{1+pR^3/3D_{\theta}}$. Thus, for high pressures this factor scales inversely with the pressure.
This flattening of the torque-curvature relation can be associated to the ovalization of the cross-sections, that become approximately elliptical as the tube is bent. The high pressure resists this effect and tends to maintain a circular cross-section. Indeed, the dominant mode of deformation of the cross-section scales as $\sin(2 \theta)$, and its magnitude follows the same functional dependence on $p$ as the non-linear term in the torque-curvature relation. Interestingly, Calladine solved the related problem of deformation of a pressurized \emph{straight} shell subject to periodic loading, and found that the effect of pressure is to repress the cross-section deformation via precisely the same functional form described above \cite{calladine1983}.

Another point of biological relevance regards the \emph{existance} of the turgor pressure in wild-type \textit{E. coli}. We note that using Equation \ref{eq:k_w_p_0}, in the absence of turgor pressure the cells would buckle at the remarkably small curvature of $0.6\% \frac{1}{R}$; in other words, the cell wall would collapse upon any minor mechanical perturbation. The turgor pressure therefore plays an important role in stabilizing the shell, though this point has been largely overlooked in the biological literature.  For the \textit{E. coli} parameters, the critical curvature for the wrinkling instability is predicted to be of the order of the inverse cell radius, allowing the cell to undergo severe mechanical deformations. This is consistent with the remarkable flexibility of growing cells to adapt to narrow microfludic constrictions \cite{wong2017}.

In summary, we have provided here an analysis of the Brazier instability in thin, pressurized tubes, and fully characterized the dramatic role of pressure in suppressing the instability. This can be naturally used as a tool for measuring turgor pressure in bacteria, and potentially other cell-walled organisms such as plants. Our results pave the way to future studies on pressurized shells. In particular, while here the analysis is performed at the level of linear incremental constitutive laws (though geometrical non-linearities are fully accounted for), it would be interesting to see how the results would carry over to the case of neo-hookean elastic models, or more elaborate models of bacterial cell walls. Furthermore, in the future it would be interesting to explore how sensitive the onset of the instability is to imperfections, as has been studied in the context of other elastic instabilities of pressurized shells, albeit with external pressure larger than the internal one, leading to implosions \cite{virot2017}\cite{paulose2013}. Finally, it would be interesting to explore -- analytically, numerically and experimentally -- the development of the onset of wrinkling into an instability, as was studied in other systems in previous works \cite{diamant2011}\cite{davidovitch2012}.

\textit{Acknowledgements:}
We thank Felix Wong, Efi Efrati and Hillel Aharony for insightful discussions. A.A. acknowledges funding from the Volkswagen Foundation, NSF CAREER Award 1752024 and MRSEC grants DMR-1420570 and DMR-2011754.

\bibliography{bib}

\end{document}


\title{Supplementary Material - Bending Instability of Rod-shaped Bacteria}

\author{Luyi Qiu}
\address{John A. Paulson School of Engineering and Applied Sciences, Harvard University, Cambridge, MA 02138, USA}

\author{John W. Hutchinson}
\address{John A. Paulson School of Engineering and Applied Sciences, Harvard University, Cambridge, MA 02138, USA}

\author{Ariel Amir}
\address{John A. Paulson School of Engineering and Applied Sciences, Harvard University, Cambridge, MA 02138, USA}


\maketitle

\section{Incremental model and corresponding energy expression} 

The incremental material response measured from the cylindrical pressurized state is taken to be linear and governed by the general orthotropic relation for stress and strain increments in the axial and circumferential directions:
\begin{equation}
    \Delta \varepsilon_z = \frac{\Delta \sigma_z - \nu_{z \theta}\Delta \sigma_{\theta}}{E_z}, \;
    \Delta \varepsilon_{\theta} = \frac{\Delta \sigma_{\theta} - \nu_{\theta z}\Delta \sigma_z}{E_{\theta}}
    \label{eq:incremental_stress_strain}
\end{equation}
with $\nu_{z \theta}/E_z = \nu_{\theta z} / E_{\theta}$.
The incremental moduli and contraction ratios should be evaluated
in the cylindrically pressurized state.
For a tube under appreciable pressure, such as the \textit{E. coli}, the cylindrical deformation due to $p$ may involve nonlinear elastic deformation, depending on
its constitutive model.
The relations between the increments of resultant membrane stresses and shell wall bending moments and the increments of stretching strains and curvatures which follow from Equation \ref{eq:incremental_stress_strain} are
\begin{equation}
    \begin{aligned}
       & \Delta N_z = S_z(\Delta \varepsilon_z + \nu_{\theta z}\Delta \varepsilon_{\theta}),
        \Delta N_{\theta} = S_{\theta}(\Delta \varepsilon_{\theta} + \nu_{z\theta}\Delta \varepsilon_z) \\
        & \Delta M_z = D_z(\Delta K_z + \nu_{\theta z}\Delta K_{\theta}),
        \Delta M_{\theta} = D_{\theta}(\Delta K_{\theta} + \nu_{z \theta}\Delta K_z)\\
    \end{aligned}
    \label{eq:incremental_N_M}
\end{equation}
where $S_z= E_z t /(1-\nu_{\theta z}\nu_{z \theta})$, $S_{\theta}= E_{\theta} t /(1-\nu_{\theta z}\nu_{z \theta})$,
$D_z=S_z t^2 /12$ and $D_{\theta}=S_{\theta} t^2 /12$.

The two contributions to the elastic energy that Brazier chose to characterize this problem are the circumferential bending energy and the axial stretching energy. 
The elastic energy change from the cylindrical state per unit length in this Brazier approximation is:
\begin{equation}
  \Delta \Phi_E =
  \Delta \Phi_s + \Delta \Phi_b
  = \int_0^{2\pi R}  \left( N_z^0 \Delta \varepsilon_z + \frac{1}{2}\Delta N_z \Delta \varepsilon_z \right)ds
   + \int_0^{2\pi R}  \left( \frac{1}{2}\Delta M_{\theta} \Delta K_{\theta} \right)ds\;.
\end{equation}

\begin{figure*}[!h]
    \centering
    \includegraphics[width=\columnwidth]{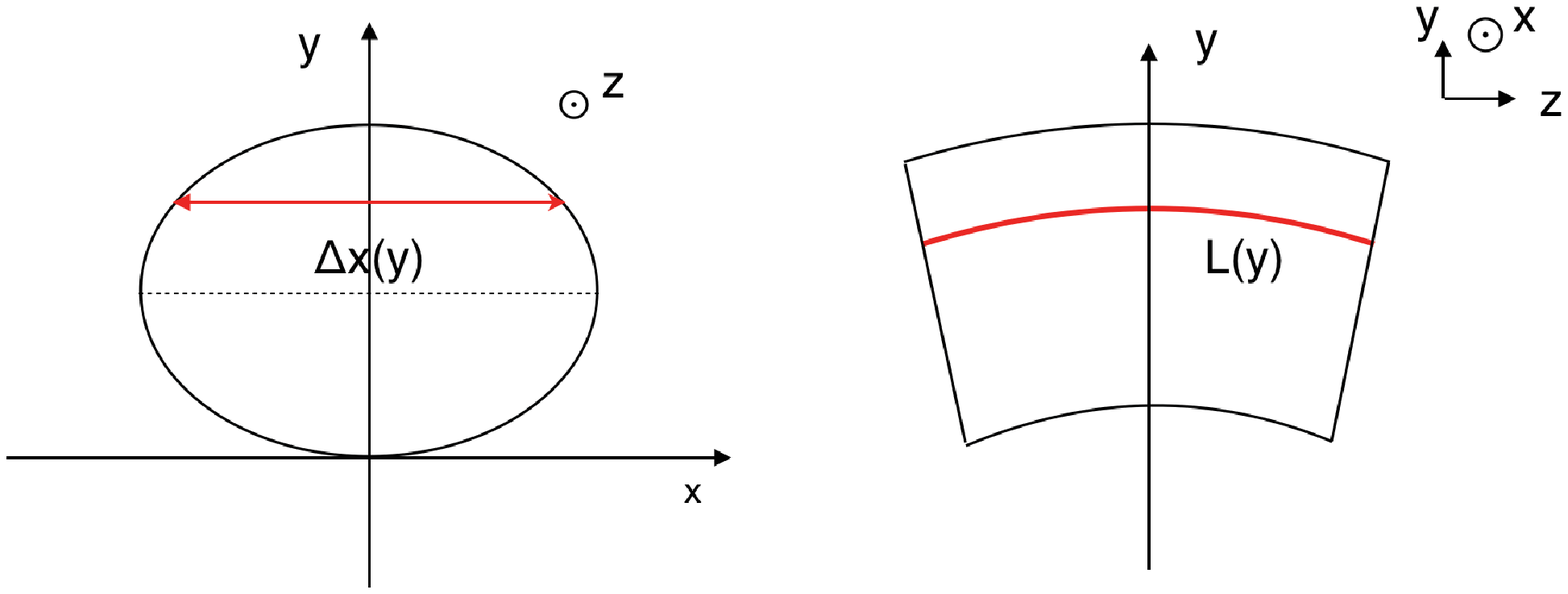}
    \caption{\textbf{Geometry notation for volume calculation.}
    \textbf{Left:} Cross section of the bent tube.
    \textbf{Right:} Side view of the bent tube.}
    \label{fig:volume}
\end{figure*}

As explained in the main text, the axial strain change in the bending step varies linearly across the cross-section according to $\Delta \varepsilon_z = \Delta \varepsilon_0 + \kappa y $, where $ \Delta \varepsilon_0$ is the change of axial strain at $y = 0$ (the bottom of the cross-section, see Figure \ref{fig:volume}) and $\kappa$ is the overall imposed curvature along the bottom of the shell at $y = 0$. Using the expressions $\Delta \varepsilon_z = \Delta \varepsilon_0 + \kappa y$, $\Delta \varepsilon_{\theta}=0$ and Equation \ref{eq:incremental_N_M}, we have:
\begin{equation}
  \Delta \Phi_s = p \pi R^2\left(\Delta \varepsilon_0+\frac{\kappa}{2\pi}\int_0^{2\pi} yd\mu \right)
  + \frac{1}{2} \int^{2\pi}_{0}  S_z(\Delta \varepsilon_0 + \kappa y)^2 R d\mu\;.
\end{equation}
For the circumferential bending energy, $\Delta K_{\theta}= d \phi/ ds $ is the circumferential curvature change and $\Delta M_{\theta}$ is the associated change of circumferential bending moment per length carried by the shell. So we have
\begin{equation}
    \Delta K_{\theta} =\frac{d \beta}{d s} - \frac{1}{R}
    = \frac{1}{R}\frac{d(\phi+\mu)}{ d\mu } - \frac{1}{R}
    = \frac{1}{R}\frac{d \phi}{ d\mu } \;.
\end{equation}
Also neglecting the curvature change in the longitudinal direction $\Delta K_z $ and using Equation \ref{eq:incremental_N_M} we have:
\begin{equation}
    \Delta \Phi_b= \frac{1}{2}\int_0^{2\pi R} \Delta M_{\theta} \Delta K_{\theta} ds
    = \frac{1}{2}\int_0^{2\pi R} D_{\theta} K_{\theta}^2 ds
    = \frac{1}{2}\int_0^{2\pi}\frac{D_{\theta}}{R^2}\left(\frac{d\phi}{d\mu}\right)^2 R d\mu\;.
\end{equation}

To calculate the tube volume change per unit length $\Delta V$, 
we define the width in the $x$ direction of the cross section as $\Delta x(y)$ and the length of the tube $L(y)$ as a function of $y$ as shown in Figure \ref{fig:volume}.
Ignoring end effects, the volume of the tube can be expressed as:
\begin{equation}
  \begin{aligned}
      V & =\int_0^{y_{max}}\Delta x(y) L(y) dy \\
      & =\int_0^{y_{max}}\Delta x L (1+\Delta \varepsilon_z) dy  \\
      & =L\int_0^{y_{max}}\Delta x (1+ \Delta \varepsilon_0 + \kappa y) dy  \\
      & =L\int_0^{\pi} 2 x (1+ \Delta \varepsilon_0 + \kappa y) \left(\frac{dy}{d\mu}\right) d\mu \\
      & =L\int_0^{2 \pi} x R (1+ \Delta \varepsilon_0 + \kappa y) \sin(\mu+\phi) d\mu \;.\\
  \end{aligned}
\end{equation}
So the tube volume change from the straight pressurized state per unit length is
\begin{equation}
  \begin{aligned}
      \Delta V & =\int_0^{2 \pi} x R (1+ \Delta \varepsilon_0 + \kappa y) \sin(\mu+\phi) d\mu -\pi R^2 \;.\\
  \end{aligned}
\end{equation}

Thus the change of energy per unit length for deformation function $\phi(\mu)$ is:
\begin{equation}\label{eq:energy_change}
  \begin{aligned}
    \Delta \Phi & = \Delta \Phi_s +\Delta \Phi_b - p \Delta V \\
    & = \frac{1}{2}\int_0^{2\pi}
    \left[\frac{D_{\theta}}{R^2}\left(\frac{d\phi}{d\mu}\right)^2 +S_z (\Delta \varepsilon_0+\kappa y)^2\right] R d\mu \\
    & +p \pi R^2\left(1+\Delta \varepsilon_0+\frac{\kappa}{2\pi}\int_0^{2\pi} yd\mu \right) \\
    & -p R ^2 \int_0^{2\pi} \frac{x}{R}(1+\Delta \varepsilon_0+\kappa y) \sin(\mu+\phi)d\mu \;. \\
  \end{aligned}
\end{equation}

\newpage
\section{typical elastic properties of \textit{E. coli} cells}
In this section we review the various elastic properties of \textit{E. coli} cells in literature. 
\renewcommand{\arraystretch}{1.5}
\begin{table}[h!] 
  \centering
  \begin{ruledtabular}
  \begin{tabular}{ c c c}
    \textbf {Properties of \textit{E. coli} cells} & \textbf {Estimated value} & \textbf {Typical value chosen for numerics} \\ 
    \hline
   Cell wall axial Young's modulus, $E_z$ & 20-30 MPa  \cite{deng2011} \cite{amir2014} \cite{yao1999} & 25 MPa\\
    Circumferential Young's modulus, $E_{\theta}$  & 50-75 MPa \cite{deng2011} \cite{lan2007} & 60 MPa\\
   Cell wall thickness, $t$ & 3-6 nm \cite{yao1999} \cite{gan2008} &  5 nm   \\
   Cell wall Poisson ratio, $\nu$ & 0.2  \cite{yao1999} & 0.2\\
   Turgor pressure, $p$ & 0.3-2 atm \cite{deng2011} \cite{koch1990} \cite{cayley2000} & 1 atm \\
   Cell cross section radius, $R$ & 0.5 $ \mu$m & 0.5 $ \mu$m \\
  \end{tabular}
\end{ruledtabular}
  \caption{\textbf{Typical elastic properties of \textit{E. coli} cells}}
  \label{table:property}
\end{table}

\section{Determination of \texorpdfstring{$\Delta \varepsilon_0$}{} }
Here we show two equivalent ways to determine the strain $\Delta \varepsilon_0=\Delta E_z(y=0)$: energy minimization with regard to $\Delta \varepsilon_0$ and the force-balance equation. 
Consider the expression for the total energy change of the system (Equation \ref{eq:energy_change}),
\begin{equation}
  \begin{aligned}
    \Phi & = \frac{1}{2}\int_0^{2\pi}
    \left[\frac{D_{\theta}}{R^2}\left(\frac{d\phi}{ds}\right)^2 +S_z (\Delta \varepsilon_0+\kappa y)^2\right]R d\mu
     +p \pi R^2\left(1+\Delta \varepsilon_0+\frac{\kappa}{2\pi}\int_0^{2\pi} yd\mu \right) \\
    & -p R^2 \int_0^{2\pi} (1+\Delta \varepsilon_0+\kappa y)(x/R)\sin(\mu+\phi)d\mu \;. \\
  \end{aligned}
\end{equation}
We define $A$ as the area of deformed cross section and $\Delta A $ as the change of cross section area from the circular state. 
\begin{equation}
    A = \int_0^{2\pi}xR\sin(\mu+\phi)d\mu ,\; \Delta A=A-\pi R^2 \;.
\end{equation}
The energy minimization with regard to $\Delta \varepsilon_0$ gives us:
\begin{equation}\label{eq:varepsilon_0_compare}
  \begin{aligned}
    0=\frac{d \Phi}{d \Delta \varepsilon_0}
    & = \frac{S_zR}{2}  \int^{2\pi}_{0} 2 (\Delta \varepsilon_0+ \kappa  y) d\mu + p \pi R^2 -p A  \;,\\
    \Delta \varepsilon_0 & =\frac{p\Delta A}{2\pi S_zR}- \frac{\kappa}{2\pi}\int_0^{2\pi} y d\mu \;. \\
  \end{aligned}
\end{equation}

We can also derive the expression for $\Delta \varepsilon_0$ through the force-balance equation.
Recall the expression of stress in $z$ direction $N_z$:
\begin{equation}
  N_z =S_z\Delta \varepsilon_z + N_{z0}=S_z\left(\Delta \varepsilon_0+\kappa  y \right)+\frac{pR}{2}\;.
\end{equation}
Since we are discussing a pressurized capped tube, the ends are pushed outward
by inner pressure while held by stress along $z$ direction in the tube. Force
balance in the $z$-direction requires:
\begin{equation}
  pA = \int_0^{2\pi R} N_z ds = \int_0^{2\pi} S_z\left(\Delta \varepsilon_0+\kappa  y\right) R d\mu + p\pi R^2\;,
\end{equation}
where $p$ is the inner pressure, $A$ is the area of deformed cross section.
Thus the expression for $\Delta \varepsilon_0$ is,
\begin{equation}
  \Delta \varepsilon_0=\frac{p\Delta A}{2\pi S_zR}- \frac{\kappa}{2\pi}\int_0^{2\pi} y d\mu \;,
\end{equation}
which agrees with equation \ref{eq:varepsilon_0_compare}.

\section{Calculation of Deformation Coefficients}

We begin with the energy expression (Equation \ref{eq:energy_change}):
\begin{equation}
  \begin{aligned}
    \Delta \Phi & = \frac{1}{2}\int_0^{2\pi}
    \left[\frac{D_{\theta}}{R^2}\left(\frac{d\phi}{d\mu}\right)^2 +S_z (\Delta \varepsilon_0+\kappa y)^2\right] R d\mu 
    + p \pi R^2\left(1+\Delta \varepsilon_0+\frac{\kappa}{2\pi}\int_0^{2\pi} yd\mu \right)  \\
    & -p R^2 \int_0^{2\pi} (1 + \Delta \varepsilon_0+\kappa y)(x/R)\sin(\mu+\phi)d\mu\;, \; \mbox{with} \; 
    \Delta \varepsilon_0 = \frac{p\Delta A}{2\pi S_zR}-\frac{\kappa }{2\pi}\int_0^{2\pi }y d\mu \;.\\
  \end{aligned}
\end{equation}

\noindent Noticing $\int_0^{2\pi} (\Delta \varepsilon_0+\kappa y) d\mu  = 2\pi \Delta \varepsilon_0 + \kappa \int_0^{2\pi} y d\mu$, we add a constant $c=-\frac{p^2 R^3}{8S_z}-p\pi R^2$ to the energy expression and get:
\begin{equation}
  \begin{aligned}
    \Phi & = \frac{1}{2}\int_0^{2\pi }
    \left[\frac{D_{\theta}}{R^2}\left(\frac{d\phi}{d\mu}\right)^2 + S_z ( \varepsilon_0+\kappa y)^2\right] R d\mu  -p R^2 \int_0^{2\pi} (1+\Delta \varepsilon_0+\kappa y)(x/R)\sin(\mu+\phi)d\mu \;, \\
  \end{aligned}
\end{equation}
in which
\begin{equation}
  \begin{aligned}
  \varepsilon_0 & = \frac{p A}{2\pi S_z R}-\frac{\kappa }{2\pi}\int_0^{2\pi }y d\mu \;, \\
  A & = \int_0^{2\pi } x \sin(\mu+\phi) R d\mu = - \int_0^{2\pi} y \cos(\mu+\phi)R d\mu \;,\\
  \end{aligned}
\end{equation}

\noindent
We discretize the function $\phi(\mu)$ using the Fourier series:
\begin{equation}
  \phi(\mu) = \frac{1}{2}a_0 +\sum_{n=1}^{N} a_n \sin(n\mu)+\sum_{n=1}^{N} b_n \cos(n\mu) \;.
\end{equation}

\noindent The system is symmetric about the y-axis,
which requires $\phi(\mu)=-\phi(2\pi-\mu)$.
So $a_0$ and $b_n$ all vanish. We also require the cross-section is closed: $x(2\pi) = x(0) =0$.
So we have
\begin{equation}\label{eq:expansion_phi}
  \begin{aligned}
      \phi(\mu) & = \sum_{n=1}^{N} a_{n} \sin(n\mu) \;, \\
      0  & = \int_0^{2\pi} \cos(\mu +\phi(\mu)) d\mu \;. \\
  \end{aligned}
\end{equation}
In section \Romannum{7} we will obtain numerical solutions using Equation \ref{eq:expansion_phi} with no restrictions on the amplitudes of $a_n$. 
Here we assume that $a_n$ are small so that we can Taylor expand 
$\sin(\mu+\phi)$ and $\cos(\mu+\phi)$
(the integrands of the $x(\mu)$ and $y(\mu)$ expressions)
about $\phi(\mu) $ and keep only the lowest order term in the Taylor Expansion.

\noindent
Using the balloon non-dimensionalization we have
\begin{equation}
    \hat{\Phi}=\frac{1}{2}\int_0^{2\pi}\left[\alpha^2\left(\frac{d\phi}{d\mu}\right)^2+\left(\varepsilon_0
    + \hat{\kappa}\frac{y}{R}\right)^2\right]d\mu
    -\hat{p}\hat{V}\;,
\end{equation}
in which
\begin{equation}
\label{eq:normalization_balloon}
    \alpha = \sqrt{\frac{D_{\theta}}{S_zR^2}}\;,
    \hat{\Phi} = \frac{ \Phi}{S_z R}\;, \;
    \hat{p} = \frac{pR}{S_z} \;, \;
    \hat{\kappa} = \kappa R \;, \;
    \hat{V} =\frac{ \Delta V+ \pi R^2}{R^2} = \int_0^{2\pi} (1 + \Delta \varepsilon_0+\kappa y)(x/R)\sin(\mu+\phi)d\mu \;.
\end{equation}
For the shell non-dimensionalization we have
\begin{equation}
    \bar{\Phi}=\frac{1}{2}\int_0^{2\pi}\left[\left(\frac{d\phi}{d\mu}\right)^2+\left(\bar{\varepsilon_0}
    + \bar{\kappa}\frac{y}{R}\right)^2\right]d\mu
    -\bar{p}\bar{V}\;,
\end{equation}
in which
\begin{equation}
\label{eq:normalization_shell}
    \bar{\Phi}  = \frac{R}{D_{\theta}} \Phi \;,\;
    \bar{p}  = \frac{pR^3}{D_{\theta}} \;, \;
    \bar{\kappa}  = \frac{\kappa R}{\alpha} \;,\;
    \bar{\varepsilon_0} = \frac{\varepsilon_0}{\alpha}  \;, 
    \bar{V}=\hat{V} \;.
\end{equation}

\subsection{Zero pressure case}
We use shell non-dimensionalization for this subsection. 
When the pressure is zero, the energy expression is:
\begin{equation}
\label{eq:Phibar_zero_pressure}
    \bar{\Phi}=\frac{1}{2}\int_0^{2\pi}\left[\left(\frac{d\phi}{d\mu}\right)^2+\left(\bar{\varepsilon}_0
    + \bar{\kappa}\frac{y}{R}\right)^2\right]d\mu\;,
\end{equation}
We Taylor expand $\sin(\mu+\phi)$ and $\cos(\mu+\phi)$
(the integrands of the $x(\mu)$ and $y(\mu)$ expressions)
about $\phi(\mu) $ and keep only the lowest order term
in the Taylor Expansion.

\begin{equation}
  \begin{aligned}
    \frac{x(\mu)}{R} & = \int_0^{\mu} \cos(\mu^\prime + \phi(\mu^\prime)) d\mu^\prime \;, \\
                & \approx \int_0^{\mu} (\cos(\mu^\prime) - \phi(\mu^\prime)\sin(\mu^\prime) ) d\mu^\prime \;, \\
                & = \sin\mu+ \frac{a_1}{4}\left[\sin(2\mu) -2\mu \right] +\sum_{n=2}^{\infty} \left[ \frac{a_n\sin(n+1)\mu}{2(n+1)}
                -\frac{a_n\sin(n-1)\mu}{2(n-1)}\right]\;.\\
  \end{aligned}
\end{equation}

\begin{equation}
  \begin{aligned}
    \frac{y(\mu)}{R} & = \int_0^{\mu} \sin(\mu^\prime + \phi(\mu^\prime)) d\mu^\prime \;, \\
                & \approx \int_0^{\mu} (\sin(\mu^\prime) + \phi(\mu^\prime)\cos(\mu^\prime) ) d\mu^\prime \;, \\
                & = 1 -\cos\mu +\frac{a_1}{4}(1-\cos(2\mu)) +\sum_{n=2}^{\infty} \left[ \frac{a_n(1-\cos(n+1)\mu)}{2(n+1)}
                 +\frac{a_n(1-\cos(n-1)\mu)}{2(n-1)}\right]\;.\\
  \end{aligned}
\end{equation}
The boundary condition $x(2\pi)=0$ requires $a_1=0$.
So we have:
\begin{equation}
  \begin{aligned}
    \bar{\varepsilon}_0+\bar{\kappa}\frac{y}{R}
  &  = -\bar{\kappa} \left[ \left(\frac{a_2}{2}+1\right)\cos\mu + \frac{a_3}{4}\cos2\mu \right]
  -\bar{\kappa} \sum_{n=3}^{\infty}\frac{a_{n-1}+a_{n+1}}{2n} \cos n\mu  \;, \\
  \end{aligned}
\end{equation}
and thus
\begin{equation}
    \frac{2\bar{\Phi}}{\pi} = \sum_{n=2}^{\infty}n^2a_n^2
     + \bar{\kappa}^2 \sum_{n=1}^{\infty}\left(\frac{a_{n-1}+a_{n+1}}{2n}\right)^2 \;,
\end{equation}
in which we assign $a_0 = 2$ for mathematical convenience. Note that $a_0$ does not correspond to any form of deformation.
Requiring $\frac{d \Phi }{da_n}=0$, we have:
\begin{equation}
  2n^2 a_n + \bar{\kappa}^2 \left[ \frac{a_{n-2}+a_n}{2(n-1)^2}
  +\frac{a_n+a_{n+2}}{2(n+1)^2}\right] =0 \;.
\end{equation}
So far we assumed that the coefficients are small so that we can
Taylor expand the integrands of the $x$ and $y$ expression.
The even coefficients ($a_{2k}$, $k\in \mathbb{N}$)
and odd ones ($a_{2k+1}$, $k\in \mathbb{N}$) are decoupled.
Clearly setting all odd terms zero ($a_{2k+1}=0$, $k\in \mathbb{N}$) is one solution.
Also notice that in the undeformed state, $a_n=0$.
So $a_{2k+1}=0$ is a stable set of solution for the 
odd coefficients.
For the even terms, assuming $a_{2k} \gg a_{2k+2}$ ($k\in \mathbb{N}$), we have:
\begin{equation}
  \begin{aligned}
  &  2n^2 a_n + \bar{\kappa}^2 \left( \frac{a_{n-2}}{2(n-1)^2}\right) =0 \;, \\
  &  a_n = - \frac{\bar{\kappa}^2}{4(n-1)^2n^2}a_{n-2}  \;. \\
  \end{aligned}
\end{equation}

\noindent 
When $n\geq 4$, the reduction ratio $ \frac{1}{4(n-1)^2n^2}\leq \frac{1}{576}$ if $\bar{\kappa}$ is of order unity. 
As we show in the main text, the buckling curvature under zero pressure
$\bar{\kappa_w} \approx 1.320$ is of order unity.
Thus we know the assumption $a_{2k} \gg a_{2k+2}$ holds. 
Furthermore, note that $|a_2| = \frac{\bar{\kappa}^2}{8} \gg |a_n| (n\neq2)$ is the dominant coefficient.

\subsection{Positive pressure case}
We use the balloon non-dimensionalization for this derivation. 
We want to minimize the dimensionless energy $\hat{\Phi}$
as a function of the rotation angle $\phi(\mu)$:
\begin{equation}
    \hat{\Phi}=\frac{1}{2}\int_0^{2\pi}\left[\alpha^2\left(\frac{d\phi}{d\mu}\right)^2+\left(\varepsilon_0
    + \hat{\kappa}\frac{y}{R}\right)^2\right]d\mu
    -\hat{p}\hat{V}\;,
\end{equation}
in which
\begin{equation}
\label{eq:normalization_balloon}
\begin{aligned}
    \alpha & = \sqrt{\frac{D_{\theta}}{S_zR^2}}\;,
    \hat{\Phi} = \frac{ \Phi}{S_z R}\;, \;
    \hat{p} = \frac{pR}{S_z} \;, \;
    \hat{\kappa} = \kappa R \;, \; \\
    \varepsilon_0 & = \frac{p A}{2\pi S_z R}-\frac{\kappa }{2\pi}\int_0^{2\pi }y d\mu \;, \; A = \int_0^{2\pi } x \sin(\mu+\phi) R d\mu = - \int_0^{2\pi} y \cos(\mu+\phi)R d\mu \;,\\
    \hat{V} & = \int_0^{2\pi} (1 + \Delta \varepsilon_0+\kappa y)(x/R)\sin(\mu+\phi)d\mu \;.
    \end{aligned}
\end{equation}

\noindent
As before, we Taylor expand $\sin(\mu+\phi)$ and $\cos(\mu+\phi)$
about $a_2$:
\begin{equation}\label{eq:y_positive_pressure}
  \begin{aligned}
    \frac{y(\mu)}{R} & = \int_0^{\mu} \sin(\mu^\prime + \phi(\mu^\prime)) d\mu^\prime \;,\\
                & \approx \int_0^{\mu} \left(
                \sin\mu^\prime + a_2 \cos\mu^\prime \sin2\mu^\prime
               - \frac{a_2^2}{2}\sin\mu^\prime\sin^2 2\mu^\prime \right) d\mu^\prime + O(a_2^3) \;,\\
                & = (1 +\frac{a_2}{2}-\frac{a_2^2}{4})(1-\cos\mu)
                +(\frac{a_2}{6}-\frac{a_2^2}{24})(1-\cos3\mu) +\frac{a_2^2}{40}(1-\cos 5\mu)+O(a_2^3)\;,\\
  \end{aligned}
\end{equation}
leading to:
\begin{equation}\label{eq:A_positive_pressure}
  \begin{aligned}
    \frac{A}{R^2} & = -\int_0^{2\pi}\frac{y}{R}\cos(\mu+\phi)d\mu \;, \\
                & \approx - \int_0^{2\pi}
                \left[\left(1 +\frac{a_2}{2}-\frac{a_2^2}{4}\right)(1-\cos\mu)
                +\left(\frac{a_2}{6}-\frac{a_2^2}{24}\right)(1-\cos3\mu)
                +\frac{a_2^2}{40}(1-\cos 5\mu)+O(a_2^3)
                 \right]  \\
              &   \left[ \left(1-\frac{a_2}{2}-\frac{a_2^2}{4}\right)\cos\mu + \left(\frac{a_2}{2}+\frac{a_2^2}{8}\right)\cos3\mu +\frac{a_2^2}{8}\cos5\mu +O(a_2^3) \right]  d\mu\\
  \bar{A} & =\frac{A}{\pi R^2}  =\left(1+\frac{a_2}{2}-\frac{a_2^2}{4}\right)\left(1-\frac{a_2}{2}-\frac{a_2^2}{4}\right)
      +\left(\frac{a_2}{6}-\frac{a_2^2}{24}\right)\left(\frac{a_2}{2}+\frac{a_2^2}{8}\right)
      +\frac{a_2^2}{40} \cdot \frac{a_2^2}{8} +O(a_2^3) \;,\\
             & = 1-\frac{2a_2^2}{3}+ O(a_2^3) \;.
  \end{aligned}
\end{equation}
So we have,
\begin{equation}
  \begin{aligned}
    \epsilon_0+\hat{\kappa}\frac{y}{R}
  &  = \frac{\hat{p}}{2} \left(1-\frac{2a_2^2}{3}\right)
  -\left(1 +\frac{a_2}{2}-\frac{a_2^2}{4}\right)\hat{\kappa}\cos\mu
  -\left(\frac{a_2}{6}-\frac{a_2^2}{24}\right)\hat{\kappa}\cos3\mu
  -\frac{a_2^2}{40}\hat{\kappa}\cos 5\mu +O(a_2^3) \;,\\
  \end{aligned}
\end{equation}
thus
\begin{equation}
  \begin{aligned}
    \int_0^{2\pi} (\epsilon_0+\hat{\kappa}\frac{y}{R})^2 d\mu
    = \frac{\pi\hat{p}^2}{2} \left(1-\frac{4a_2^2}{3}\right)
  + \hat{\kappa}^2 \pi \left(1+a_2 - \frac{2}{9}a_2^2 \right)+ O(a_2^3) \;. \\
  \end{aligned}
\end{equation}

For the zero pressure case, we showed in the previous subsection that the dominant coefficient is $a_2$, corresponding to $\phi(\mu)=-\frac{\bar{\kappa}^2}{8}\sin (2\mu)$. Here, we will make the assumption that this is also the case for the pressurized case. This assumption is corroborated by the numerical results shown in Figure 2 in the main text which do not make this simplifying assumption in the Fourier series expansion. 
The system is then symmetric about both $y$ axis and horizontal centerline 
(the dashed line in Figure \ref{fig:volume}). 
Thus we can simplify the $\bar{V}$ expression as:

\begin{equation}
  \begin{aligned}
    \hat{V} & = \left(1+\Delta \varepsilon_0+\kappa \frac{1}{2\pi}\int_0^{2\pi}yd\mu\right)
    \int_0^{2\pi}(x/R)\sin(\mu+\phi)d\mu \;, \\
    & = \left(1+\frac{p \Delta A}{2\pi S_z R} \right) \frac{A}{R^2}
     = \left[1+ \frac{\hat{p}}{2} \left(- \frac{2a_2^2}{3}\right)\right] \pi \left(1- \frac{2a_2^2}{3}\right) + O(a_2^3)\;. \\
  \end{aligned}
\end{equation}

The energy expression thus takes the form:
\begin{equation}\label{eq:energy}
  \begin{aligned}
  \hat{\Phi} & = 2 \alpha^2 a_2^2\pi
   +\frac{\pi\hat{p}^2}{4} \left(1-\frac{4a_2^2}{3}\right)
   +\frac{\hat{\kappa}^2\pi}{2} \left(1+a_2 - \frac{2}{9}a_2^2 \right)
   -\pi \hat{p} \left[1+ \frac{\hat{p}}{2} \left(- \frac{2a_2^2}{3}\right)\right]  \left(1- \frac{2a_2^2}{3}\right)  + O(a_2^3) \;, \\
  & = 2 \alpha^2 a_2^2\pi
   +\frac{\hat{\kappa}^2\pi}{2} \left(1+a_2 - \frac{2}{9}a_2^2 \right) 
   +\frac{2a_2^2}{3} \pi \hat{p} -\pi \hat{p} + \frac{\pi}{4}\hat{p}^2 + O(a_2^3) \;. \\
  \end{aligned}
\end{equation}
By requiring $\frac{d\Phi}{da_2}=0$, we have:
\begin{equation}\label{eq:energy_min}
  0 = 4 \alpha^2 a_2 +\frac{\hat{\kappa}^2}{2} \left(1 - \frac{4}{9}a_2\right)
  +\frac{4a_2}{3}\hat{p} +O(a_2^2) \;.
\end{equation}
For each term we just keep the lowest order term of $a_2$, Thus
\begin{equation}
  4 \alpha^2 a_2 + \frac{\hat{\kappa}^2}{2}
  +\frac{4a_2}{3}\hat{p} +O(a_2^2) \approx 0 \;.
\end{equation}
Finally we obtain:
\begin{equation}\label{eq:a2}
  a_2 =-\frac{1}{\alpha^2+\hat{p}/3}\frac{\hat{\kappa}^2}{8} 
  =-\frac{1}{1+\bar{p}/3}\frac{\bar{\kappa}^2}{8}\;.
\end{equation}

\section{Torque curvature relationship}
\noindent From the dimensionless energy expression (Eqn \ref{eq:energy}) and the expression of $a_2$
(Eqn \ref{eq:a2}), we have
\begin{equation}
  \begin{aligned}
    \hat{\Phi} & = 2\pi \alpha^2 \zeta^2\hat{\kappa}^4
     +\frac{\pi \hat{\kappa}^2}{2} \left(1+\zeta\hat{\kappa}^2 
     \right)
     + \frac{2\pi\zeta^2\hat{p}}{3}\hat{\kappa}^4
     -\hat{p}\pi +\frac{\pi}{4}\hat{p}+ O(\hat{\kappa}^6) \;,\\
  \end{aligned}
\end{equation}
in which
\begin{equation}
  \zeta = -\frac{1}{8(\alpha^2+\hat{p}/3)}\; , \; a_2=\zeta \hat{\kappa}^2 \;.
\end{equation}

\noindent We can calculate the torque by
$M=\frac{\partial \Phi}{\partial \kappa} $, in which $\Phi$ is the energy of the system
per unit length:
\begin{equation}
  \begin{aligned}
    M & = \frac{\partial \Delta \Phi}{\partial \kappa} 
     = S_z R^2 \frac{\partial \hat{\Phi}}{\partial \hat{\kappa}}  \\
    & = S_z R^2 \pi \hat{\kappa} 
    + S_z R^2 \pi \zeta \hat{\kappa}^3 
    +O(\hat{\kappa}^5) \\
    & \approx S_z R^2 \pi \left[\hat{\kappa} -\frac{\hat{\kappa}^3}{8(\alpha^2+\hat{p}/3)}  \right] \\
    & = S_z R^2 \pi \left[\kappa R -\frac{\kappa^3 R^3}{8\left(\alpha^2+\frac{pR}{3S_z}\right)}  \right] \;.
  \end{aligned}
\end{equation}
Since $\hat{M}  = \frac{M}{S_z R^2}$, we have:
\begin{equation} \label{eq:torque_balloon}
  \hat{M} = \pi \left[\hat{\kappa} -\frac{\hat{\kappa}^3}{8(\alpha^2+\hat{p}/3)}  \right] \;.
\end{equation}
In shell non-dimensionalization, this relation can be expressed as
\begin{equation}\label{eq:torque_shell}
  \bar{M} = \pi \left[\bar{\kappa} -\frac{\bar{\kappa}^3}{8(1+\bar{p}/3)}  \right] \;.
\end{equation}

\section{Wrinkling critical curvature under zero and large pressure}
The critical wrinkling curvature satisfies the following equation:
\begin{equation}
  N_z = \frac{pR}{2} + S_z (\Delta \varepsilon_0+\kappa y(\mu)) = -\sigma_c t = - \frac{2}{\rho(\mu)}\sqrt{D_z S_{\theta}} 
  =- \frac{2}{\rho(\mu)}\sqrt{D_{\theta}S_z} \;,
\end{equation}
in which $\frac{1}{\rho}$ is the circumferential curvature and
$\Delta \varepsilon_0$ is the change of strain at position $y = 0$:
\begin{equation}
    \begin{aligned}
     \frac{1}{\rho} &  = \frac{1}{R} \frac{d\beta}{d\mu} = \frac{1}{R}\left( 1 +\frac{d\phi}{d\mu} \right)
     = \frac{1}{R}(1+2a_2\cos{2\mu})
     = \frac{1}{R}\left(1-\frac{\hat{\kappa}^2}{4(\alpha^2+\hat{p}/3)}\cos{2\mu}\right) \;,\\
  \Delta \varepsilon_0  & =\frac{p\Delta A}{2\pi S_zR}-\kappa  \frac{1}{2\pi}\int_0^{2\pi} y d\mu \;.\\
    \end{aligned}
\end{equation}
Clearly $y = 0$ ($\mu=0$) is the place with the largest compressive stress $N_z$. It is also the place with the largest local circumferential curvature $\rho$, thus the smallest $\sigma_c$. Therefore the wrinkling instability will happen at $y = 0$ ($\mu=0$).
This leads to the following equation for the critical wrinkling point:
\begin{equation}
  \begin{aligned}
   \frac{pR}{2} + S_z \Delta \varepsilon_0 & = - \frac{2}{\rho(y=0)}\sqrt{D_{\theta}S_z} \;,\\
  \Rightarrow  \frac{pA}{2\pi S_z R} - \kappa \frac{1}{2\pi}\int_0^{2\pi} y d\mu
  & = -2\sqrt{\frac{D_{\theta}}{S_z R^2}}(1+2a_2) \;,\\
  \end{aligned}
\end{equation}
When we use the ansatz $\phi(\mu)=a_2\sin(2\mu)$, we have expressions for $y(\mu)$ (Equation \ref{eq:y_positive_pressure}) and $A$ (Equation \ref{eq:A_positive_pressure}) leading to: 
\begin{equation}
  \begin{aligned}
      \frac{1}{2\pi}\int_0^{2\pi} y d\mu
     & = 1 +\frac{2 a_2}{3} +O(a_2^2)\;,\\
      A & = \pi R^2 \left[1-\frac{2a_2^2}{3}+ O(a_2^3) \right] \;.\\
  \end{aligned}
\end{equation}
So we have 
\begin{equation}\label{eq:k_w_original}
    8\left(\alpha^2+\frac{\hat{p}}{3}\right)
    \left(\frac{\hat{p}}{2}+2\alpha-\hat{\kappa}_w\right)
    +\left(\frac{2\hat{\kappa}_w}{3}-4\alpha\right)\hat{\kappa}_w^2=0
\end{equation}
When the pressure $p=0$, we have a simple equation for the critical wrinkling curvature $\kappa_w$:
\begin{equation}
    \left(\frac{\kappa_w R}{\alpha}\right)
    - \frac{1}{12}\left(\frac{\kappa_w R}{\alpha}\right)^3
    =2-\frac{1}{2} \left(\frac{\kappa_w R}{\alpha}\right)^2 \;.
\end{equation}
And the smallest positive solution for $\kappa_w$ is
\begin{equation}
     \bar{\kappa}_w = \frac{\kappa_w R}{\alpha} = 1.320 \;.
\end{equation}
When the pressure is in the “balloon range” and the tube is thin ($\alpha \ll 1$) such that $\alpha$ is negligible in Equation \ref{eq:k_w_original}, we have:
\begin{equation} \label{eq:k_w_small_alpha}
    \hat{\kappa}_w^3 - 4 \hat{p}\hat{\kappa}_w
    +2\hat{p}^2 =0 \;.
\end{equation}
This can be rewritten as 
\begin{equation} \label{eq:k_w_small_alpha}
    \hat{\kappa}_w= \frac{\hat{p}}{2} + \frac{\hat{\kappa}_w^3}{4\hat{p}} \;.
\end{equation}
For $\hat{p}<1$, we can expand $\hat{\kappa}_w$ in series of $\hat{p}$ and 
the first order approximation is 
\begin{equation}\label{eq:k_w_p/2}
    \hat{\kappa}_{w0} \approx \frac{\hat{p}}{2} \;.
\end{equation}
We note that for \textit{E. coli}, $\hat{p}\approx 0.4$ (Table \ref{table:property}).

\section{Numerical Energy minimization}
We use Lagrange multipliers to minimize the total energy change $\Delta \Phi$ (Equation \ref{eq:energy_change}) under the boundary constraint:
\begin{equation}
    \int_0^{2\pi} \cos(\mu +\phi(\mu)) d\mu =0 \;.
\end{equation}
We define the function $F$ as 
\begin{equation}
    F= \Delta \Phi + \lambda \int_0^{2\pi} \cos(\mu +\phi(\mu)) d\mu \;.
\end{equation}
For optimization under constraints, we wish to find a point $(a_1, a_2, ..., \lambda)$ 
such that 
\begin{equation}
  \begin{aligned}
    \nabla_{\vec{a}} F & = \nabla_{\vec{a}}(\Delta \Phi) +\lambda \nabla_{\vec{a}} \left(\int_0^{2\pi} \cos(\mu +\phi(\mu)) d\mu \right) = 0 \;, \\
    \nabla_{\lambda} F & = \int_0^{2\pi} \cos(\mu +\phi(\mu)) d\mu = 0 \;.\\
  \end{aligned}
\end{equation}
This is equivalent to
\begin{equation}
  \nabla F (\vec{a},\lambda) = 0 \;.
\end{equation}
We use Newton's method to find the point satisfying $\nabla F(\vec{a},\lambda) = 0 $ and allow up to four terms $(a_1,a_2,a_3,a_4)$ in the Fourier expansion. We verify that $a_4$ is already small $a_4 \ll a_2$. So we expect the result to be unchanged with more coefficients in the Taylor expansion. 
Please refer to \url{https://github.com/LuyiQiu/buckling_thin_shell} for the code and more details.

\section{The instability point for different systems}

\begin{figure}[!h]
    \centering
    \includegraphics[width=0.49\columnwidth]{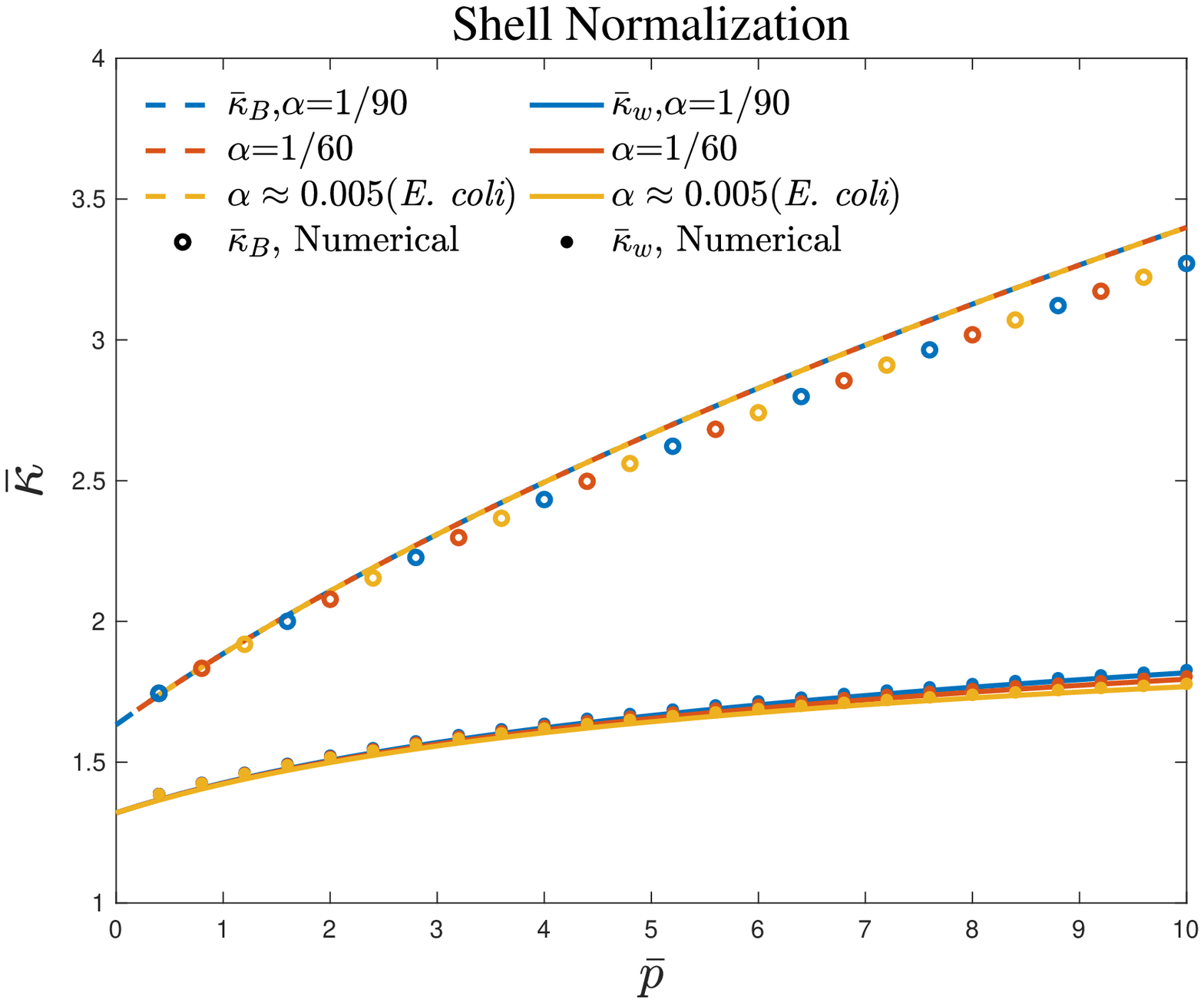}
    \includegraphics[width=0.49\columnwidth]{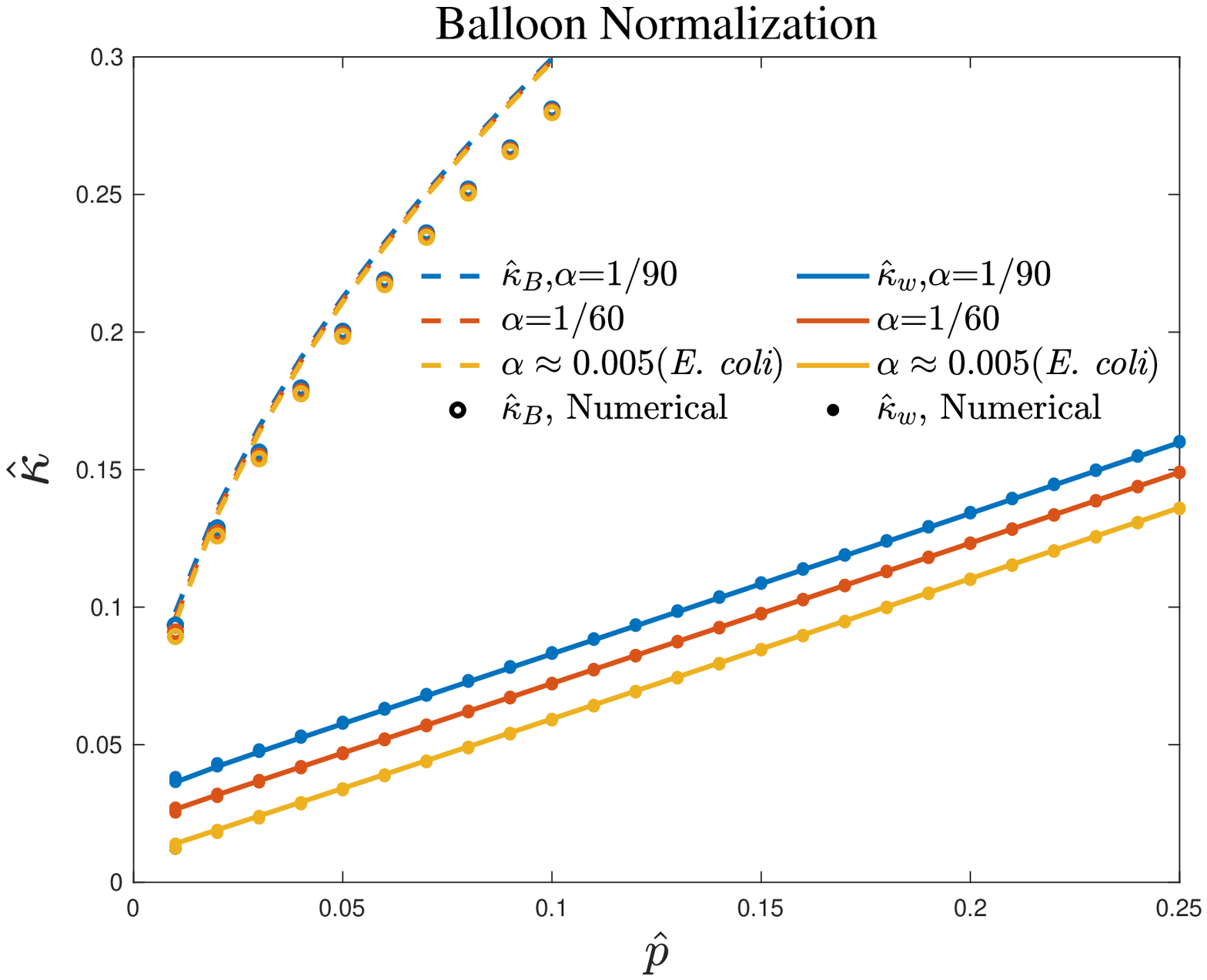}
    \caption{\textbf{Comparison of analytical and numerical results for the critical curvatures for Shell (Left) and Balloon (Right) normalizations.} 
    }
    \label{fig:critical_curvature}
\end{figure}

The dependence of $\kappa_w$ (Equation \ref{eq:k_w_original}) and the $\kappa_B$ (from the maximum of Equations \ref{eq:torque_balloon}, \ref{eq:torque_shell}) on pressure $p$ and $\alpha$ is shown in Figure \ref{fig:critical_curvature} for the two normalizations. 
On each $\kappa-p$ curve are solid or open dots present the corresponding instability points computed numerically. 
In the shell regime in Figure \ref{fig:critical_curvature}(A), the parameter $\alpha$ has no effect on $\kappa_B$ and a small influence on $\kappa_w$. In the balloon regime in Figure \ref{fig:critical_curvature}(B), $\alpha$ has very little influence on $\kappa_B$ and a modest influence on $\kappa_w$.


\newpage
\section{Turgor Pressure measurement experiment}
\begin{figure}[!h]
    \centering
    \includegraphics[width=0.5\columnwidth]{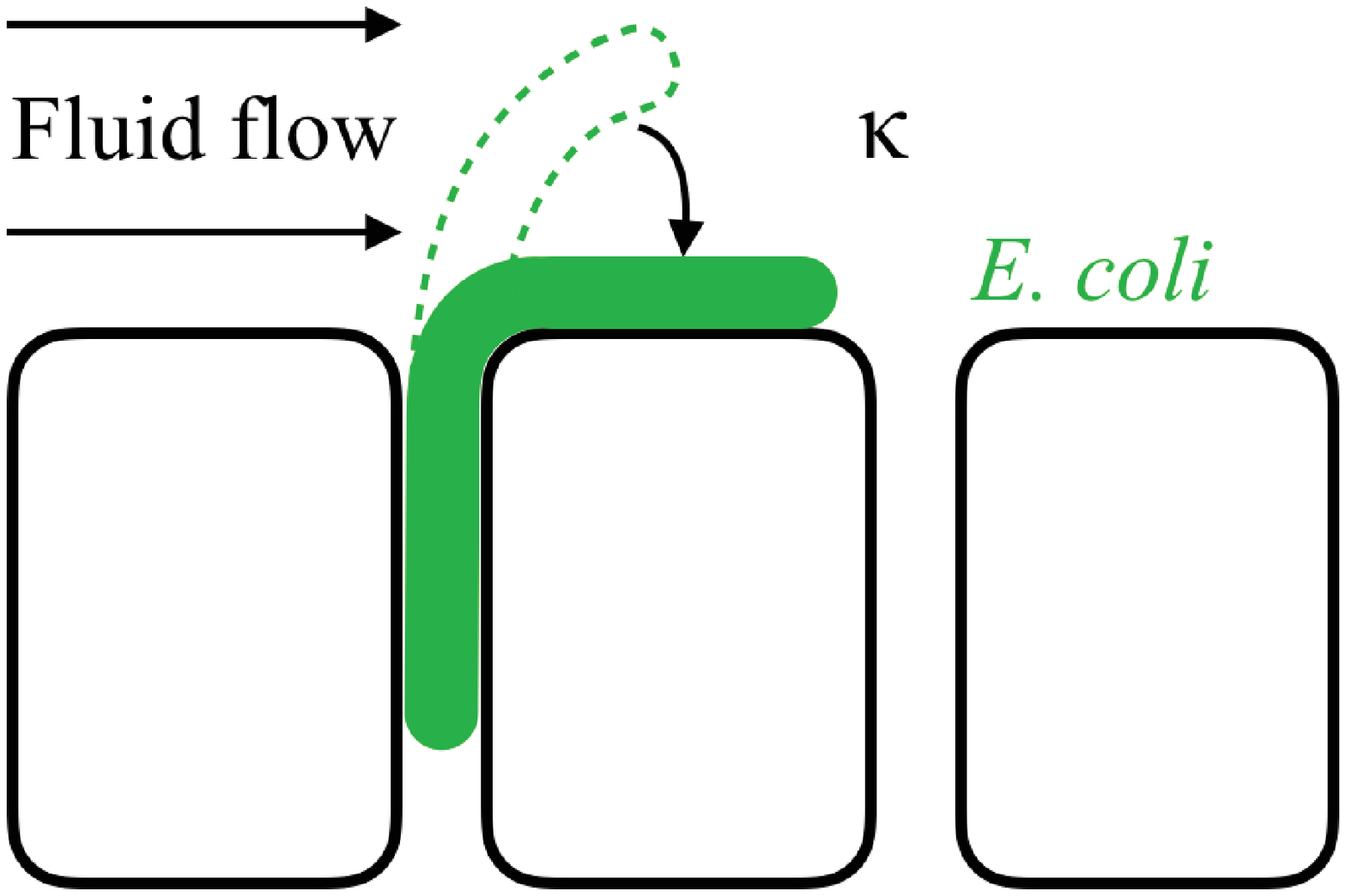}
    \caption{\textbf{An illustration of a turgor pressure measurement experiment.} }
    \label{fig:experiment}
\end{figure}
    
    We now propose an experiment relying on the theoretical results derived here and enabling the inference of the turgor pressure in live cells. 
    First, one would grow filamentous \textit{E. coli} cells in a mother machine.
    By applying hyperosmotic shock during flow, one may instantaneously lower the turgor pressure and the cell would buckle when the critical buckling curvature is reached 
    (for the particular torque exerted by the flow). 
  We can back out the turgor pressure from the lowest osmolarity in which the cell is observed to buckle.
  To see this, we define $c_1<c_2$ as the two successive medium concentration for which the cell changes from non-buckling to buckling. 
  $c_1$ would give us a lower bound of the turgor pressure according to Equation \ref{eq:k_w_p/2}:
  \begin{equation}
      \kappa_{ex1}<\kappa_w(p)=\frac{p(c_1)}{2S_z} \;,
  \end{equation}
  in which $S_z= E_z t /(1-\nu_{\theta z}\nu_{z \theta})$, $\kappa_{ex1}$ is the curvature of the cell measured from the experiment 
  (at the tip of the microfluidic channel, where it is maximal) and $p(c)=p_0-\Delta p(c)$ is the turgor pressure after the osmotic shock ($p_0$ is the turgor pressure before the osmotic shock). The osmolarity can be calculated by the Morse equation $\Delta p = RT(C_{in}-C_{out})$. 
  So we have
  \begin{equation}
      p_0 > 2 S_z \kappa_{ex1} + \Delta p(c_1)  \;.
  \end{equation}
Similarly, we can calculate an upper bound of $p_0$ from $c_2$, leading to: 
\begin{equation}
      2 S_z \kappa_{ex1} + \Delta p(c_2) > p_0 > 2 S_z \kappa_{ex1} + \Delta p(c_1) \;.
  \end{equation}
  The accuracy of the measurement can be increased by reducing the gap between successive solute concentrations used in the experiment.
  Using a continuous change in the concentration by the appropriate design of the microfluidic device would allow one to directly measure the critical curvature and turgor pressure.
\bibliography{bib}